\documentclass[lettersize,journal]{IEEEtran}
\usepackage{amsmath,amsfonts}
\usepackage{algorithmic}
\usepackage{array}
\usepackage[caption=false,font=normalsize,labelfont=sf,textfont=sf]{subfig}
\usepackage{textcomp}
\usepackage{stfloats}
\usepackage{url}
\usepackage{verbatim}
\usepackage{graphicx}

\usepackage{hyperref}
\usepackage{cleveref}
\usepackage{bbm}
\usepackage{multirow}
\usepackage{pifont}

\usepackage{graphicx}
\usepackage{caption}
\usepackage{lipsum}
\usepackage{booktabs} 
\usepackage{xcolor}
\crefname{figure}{Fig.}{Figs.}
\Crefname{figure}{Fig.}{Figs.}
\crefname{table}{Table}{Tables}
\Crefname{table}{Table}{Tables}

\def\BibTeX{{\rm B\kern-.05em{\sc i\kern-.025em b}\kern-.08em
    T\kern-.1667em\lower.7ex\hbox{E}\kern-.125emX}}
\usepackage{balance}

\begin{document}
\title{COMMA: Coordinate-aware Modulated Mamba Network for 3D Dispersed Vessel Segmentation}
\author{Gen Shi, Hui Zhang and Jie Tian, ~\IEEEmembership{Fellow,~IEEE}
}

\markboth{Journal of \LaTeX\ Class Files,~Vol.~18, No.~9, September~2020}%
{How to Use the IEEEtran \LaTeX \ Templates}

\twocolumn[{
\renewcommand\twocolumn[1][]{#1}
\maketitle
\begin{center}
    \captionsetup{type=figure}
    \includegraphics[width=\textwidth]{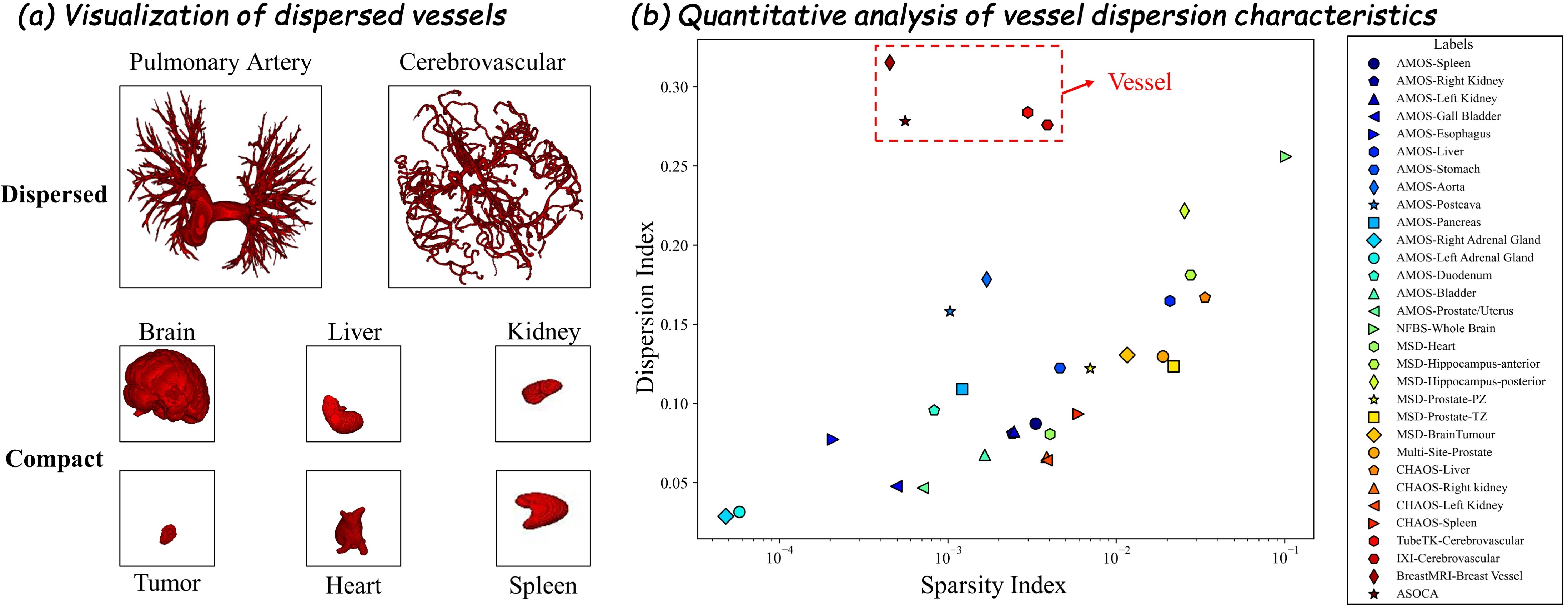}
    \captionof{figure}{Illustration of the dispersed nature of vascular structures. (a) Qualitative analysis: visualizations of dispersed vascular structures compared with other compact organs and tissues. (b) Quantitative analysis: dispersion and sparsity indices of various organs. The legend indicates 'dataset-organ' pairs.}
    \label{fig: dispersion index}
\end{center}
}]
\begingroup
\renewcommand\thefootnote{}\footnotetext{\footnotesize
This work was supported in part by the National Natural Science Foundation
of China under Grants 624B2018, 62027901, and 32371152, and the Academic Excellence Foundation of BUAA for PhD Students. (Corresponding authors: Jie Tian and Hui Zhang)}
\addtocounter{footnote}{-1}
\vspace{-0.5em}
\footnotetext{\footnotesize
Gen Shi, Hui Zhang and Jie Tian are with the School of Engineering Medicine and School of Biological Science and Medical Engineering, Beihang University, Beijing 100191, China, and also with the Key Laboratory of Big Data-Based Precision Medicine (Beihang University), Ministry of Industry and Information Technology of China, Beijing 100191, China. Jie Tian is also with CAS Key Laboratory of Molecular Imaging, Institute of Automation, Beijing, 100190, and with National Key Laboratory of Kidney Diseases, Beijing, 100853, China. (e-mail: shigen@buaa.edu.cn; hui.zhang@buaa.edu.cn; tian@ieee.org).}
\endgroup

\begin{abstract}
Accurate segmentation of 3D vascular structures is essential for various medical imaging applications. The dispersed nature of vascular structures leads to inherent spatial uncertainty and necessitates location awareness, yet most current 3D medical segmentation models rely on the patch-wise training strategy that usually loses this spatial context. In this study, we introduce the Coordinate-aware Modulated Mamba Network (COMMA) and contribute a manually labeled dataset of 570 cases, the largest publicly available 3D cerebrovascular dataset to date. COMMA leverages both entire and cropped patch data through global and local branches, ensuring robust and efficient spatial location awareness. Specifically, COMMA employs a channel-compressed Mamba (ccMamba) block to efficiently encode full-resolution image data, capturing long-range dependencies while optimizing computational costs. Additionally, we propose a coordinate-aware modulated (CaM) block to enhance interactions between the global and local branches, allowing the local branch to better perceive spatial information. We evaluate COMMA on six datasets, covering two imaging modalities and five types of vascular tissues. The results demonstrate COMMA's superior performance compared to state-of-the-art methods with computational efficiency, especially in segmenting small vessels. Ablation studies further highlight the importance of our proposed modules and spatial information. The code will be available at \href{https://github.com/shigen-StoneRoot/COMMA}{\textit{COMMA}}.
\end{abstract}


\section{Introduction}
\label{sec:intro}
The precise segmentation of 3D vascular structures is essential in various fields, particularly in medical imaging \cite{brugnara2023deep, wang2020augmenting, 10871930}. It serves as a fundamental step in vascular image analysis and is critical for tasks such as vascular morphology and dynamics analysis, aneurysm detection, and vascular intervention surgery \cite{tronolone2023evaluation, ueda2019deep}. 

Although foundation models have achieved impressive results in medical segmentation \cite{ma2024segment, wittmann2024vesselfm, wang2023sam}, they usually focus on commonly compact organs and tissues, leaving challenges in specialized areas like fine-grained 3D vascular segmentation. For example, MedSAM acknowledges the difficulty of segmenting vessel-like branching structures \cite{ma2024segment}. We show a preliminary evaluation in \cref{sec FM analysis}. Therefore, developing a specialized model for such a field is still important. 

Compared with the commonly compact organs, the segmentation of vascular structures presents two major challenges:

\textbf{(1) The dispersed nature of vascular structures necessitates spatial localization}. Unlike common compact organs (e.g., liver), vascular structures exhibit a highly dispersed distribution in space (see \cref{fig: dispersion index}a) with inherent spatial uncertainty \cite{gibbins2019innervation}, especially distal small vessels. We show a quantitative result in \cref{fig: dispersion index}b. The definitions of Dispersion Index and Sparsity Index are present in Appendix \cref{sec definition Dispersion Index}. While compact organs and tissues generally occupy fixed spatial locations, the spatial uncertainty of blood vessels poses a greater challenge, acquiring vascular spatial information crucial for accurate segmentation.

\begin{figure}
  \centering
  \includegraphics[width=\linewidth]{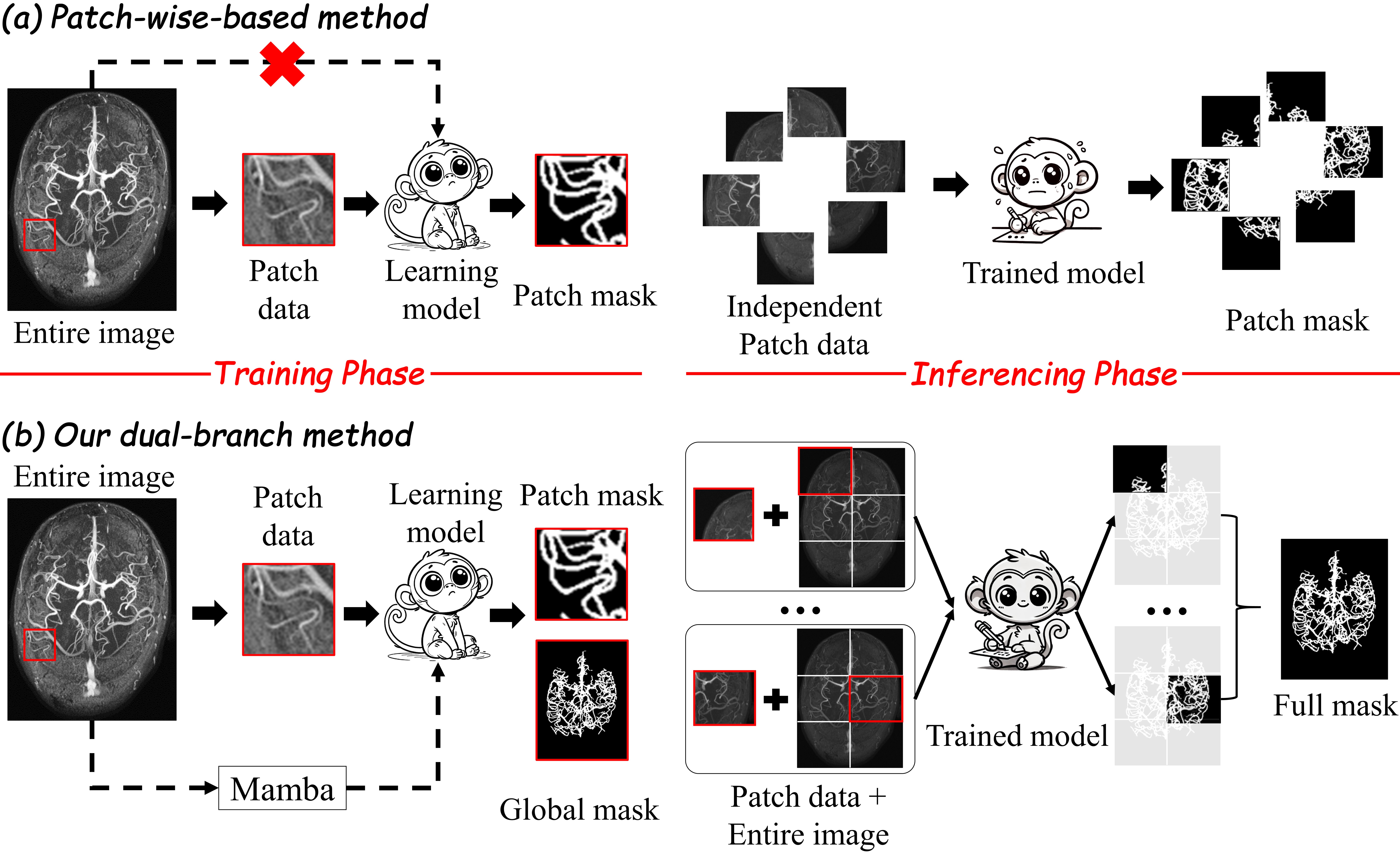}
    \caption{Illustration of the (a) patch-wise-based method compared to (b) our proposed dual-branch method.}
    \label{fig: dual branch}
\end{figure}

\textbf{(2) Failure of patch-wise strategy in capturing spatial information}. 
The patch-wise training strategy is practical for handling large 3D medical data \cite{isensee2021nnu, Basak_2023_CVPR}; however, it also results in a loss of spatial location information for cropped patches (see \cref{fig: dual branch}a). During inference, the entire image is typically divided into separate and independent patches for model input, and the spatial relationships between patches and the entire image are not preserved, which is particularly crucial in vascular structures. Thus, constructing spatial coordinate relationships that account for the entire image is essential for accurate vascular segmentation.

One challenge in using the entire 3D image is its large size. The typical 3D medical image size is (512, 512, 128). For vascular images that require high resolution, the image size may be much more large. The computation cost is high, especially for Transformer architecture \cite{azad2024medical, he2023transformers}. The computational complexity limits its application to processing the entire image data. 

Recently, the Mamba architecture has demonstrated strong success in analyzing continuous long-sequence data \cite{gu2023mamba, pmlr-v235-dao24a} and shown promise in medical image analysis \cite{xing2024segmamba, liu2024swin}. However, existing Mamba-based models primarily replace convolutional layers with Mamba layers while still operating on small patch data, failing to fully exploit Mamba's efficiency advantages. For vascular structures, Mamba presents a promising approach to processing entire images while efficiently modeling spatial coordinate relationships.

Based on the above analysis, we propose the \textbf{Co}ordinate-aware \textbf{M}odulated \textbf{Ma}mba Network (COMMA) for 3D dispersed vessel segmentation. COMMA features a dual-branch architecture that enables simultaneous processing of full 3D images and patch data (see \cref{fig: dual branch}b). Additionally, we introduce a novel coordinate-aware modulated (CaM) block to enhance spatial location awareness within the local branch. Our model was evaluated on six datasets across five distinct vessel tissues, with results demonstrating that COMMA outperforms state-of-the-art baselines. Furthermore, we contribute one of the largest 3D blood vessel datasets to date, comprising 570 cases. Our main contributions are summarized as follows: 
\begin{itemize}
    \item We propose Mamba-based dual-branch model COMMA for 3D vascular image segmentation. It has been evaluated on six datasets and consistently outperforms existing models.
    
    \item We introduce the novel CaM block and the channel-compressed Mamba (ccMamba) block. The CaM block enhances the model's spatial location awareness for patch data, while the ccMamba block reduces computational costs for processing the entire 3D image data.
    
    \item We curate and release a manually annotated dataset of 570 volumetric images based on the IXI dataset. This large dataset is intended to support researchers in designing new methods for vascular segmentation.
\end{itemize} 

\section{Related Work}
\label{sec:Related Work}
\begin{figure*}
  \centering
  \includegraphics[width=\textwidth]{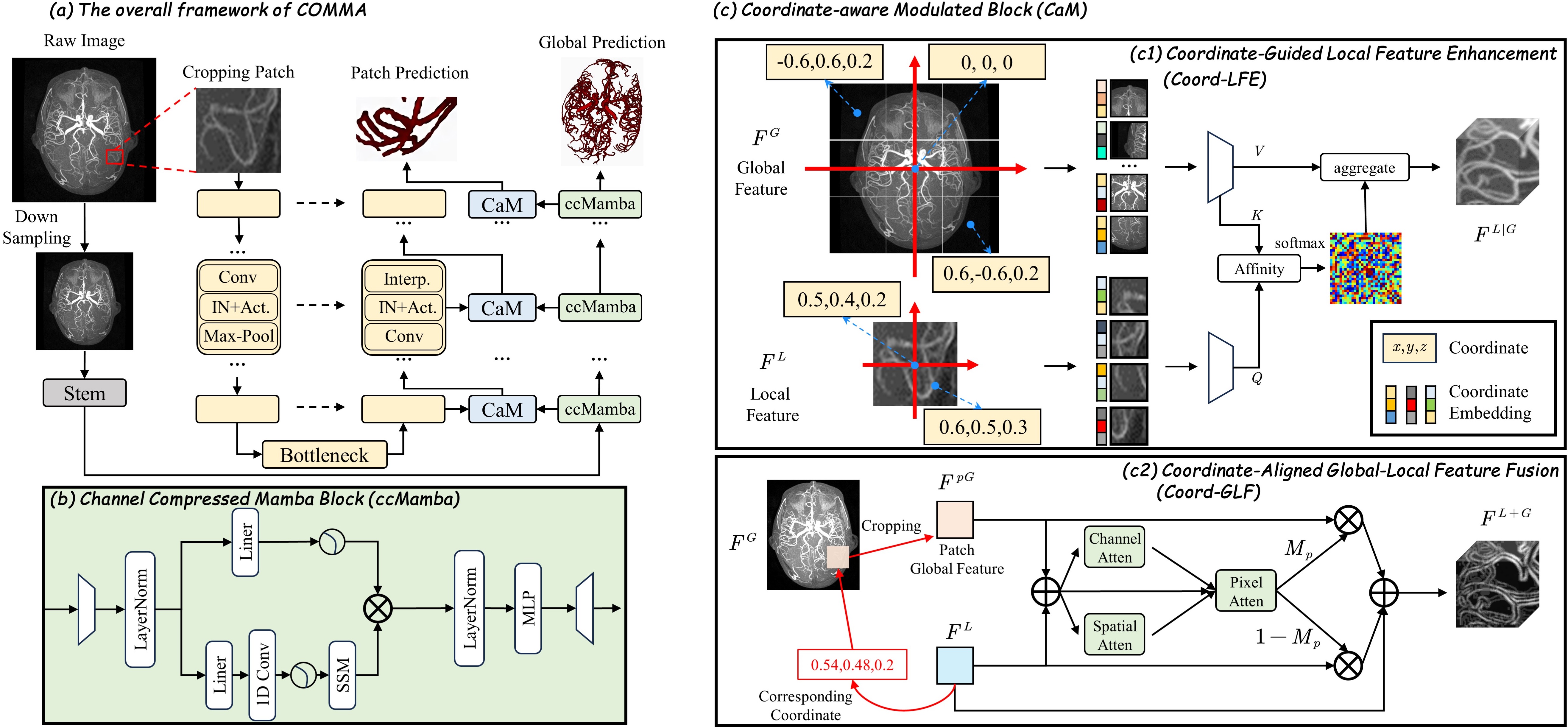}
    \caption{The overall framework of our proposed COMMA.}
    \label{fig: framework}
\end{figure*}
\subsection{3D vascular segmentation}
Conventional methods for vascular structure segmentation primarily focus on analyzing geometric characteristics. For instance, threshold-based segmentation techniques often enhance vessel structures through filters (e.g., Frangi filter) and then remove background pixels based on set thresholds \cite{akram2013multilayered, frangi1998multiscale}. Statistical model-based methods \cite{hassouna2006cerebrovascular, su20213d}, which treat pixel intensities as statistical distributions, have proven robust in merging regions based on similarity, especially effective for complex structures.

Deep learning approaches have demonstrated significant effectiveness in medical image analysis \cite{chen2023all, 11125900, 11010915, 10877692}. For example, Xia et al. introduced a network enhanced with edge information to improve the segmentation of 3D vessel-like structures \cite{xia20223d}. Qi et al. introduced DSCNet, which utilizes dynamic snake convolution and topology continuity loss for enhanced accuracy \cite{qi2023dynamic}. Transformer-based models are also emerging, capturing long-range dependencies in vessel segmentation \cite{wu2023transformer, 10183842}. Furthermore, specific loss function designs have been developed; for instance, the centerlineDice (clDice) metric considers skeletons and centerlines \cite{cldice2021}, and Shi et al. extended this with centerline boundary dice loss (cbDice) by incorporating boundary-aware factors, improving recognition of geometric details \cite{shi2024centerline}.

\subsection{Mamba for medical image segmentation}
UNet-based convolutional neural networks have played a pivotal role in medical image segmentation \cite{azad2024medical}. While CNNs excel at capturing localized information, they are less effective for modeling long-range dependencies. Transformer architectures with self-attention address this limitation \cite{han2022survey} but face a significant drawback: computational costs \cite{lin2023super, chen2021visformer}, posing challenges for large medical image data.

Recently, State Space Models, exemplified by the Mamba architecture, have gained attention \cite{gu2023mamba, pmlr-v235-dao24a}. The Mamba structure achieves a strong balance between efficiency and accuracy in analyzing long sequences. Mamba-based models have also been extended to medical image analysis \cite{bansal2024comprehensive}, with several efficient segmentation models proposed, such as SegMamba \cite{xing2024segmamba}, UMamba \cite{ma2024u}, and Swin-UMamba \cite{liu2024swin}. Besides, several recent studies have explored the application of Mamba to vessel segmentation \cite{10.1145/3696409.3700231, liu2025swin, xie2025vesselmamba}. However, current Mamba-based models typically replace the convolution or transformer layer with Mamba blocks and continue to process only patch data, limiting the full potential of the Mamba architecture for handling long sequences.

\section{The Proposed COMMA}
\subsection{Overall Framework}
The overall framework is illustrated in \cref{fig: framework}a. COMMA comprises two branches: a local branch and a global branch. Patch data, randomly cropped from the original full-resolution image, is fed into the local branch. Meanwhile, the full-resolution image is downsampled to a fixed, smaller size and input to the global branch, which uses Mamba structures (\cref{fig: framework}b). The CaM block (\cref{fig: framework}c) is central to COMMA, facilitating interactions between the decoder of the local branch and the global branch through coordinate positions. 
\subsection{The Local Branch}
The encoder part of the local branch consists of successive convolutional modules. Each convolutional module contains a 3D convolution operation, Instance Normalization, Leaky ReLU, and max-pooling. The encoder has 4 stages. The patch data $X^{L}$ is initially processed by a convolution without a pooling layer to adjust the channel dimension. In each stage, the channel dimension is doubled, while the spatial dimensions are halved. In the decoder, each convolutional module contains the same components as in the encoder, but the pooling layer is replaced with interpolation. Additionally, the output of each convolutional module is fed into the CaM block, and the output from the CaM block is used as input for the next convolutional module.
\subsection{The Global Branch}
\label{sec: The Global Branch}
The full-resolution image size is usually variable, so it is first resized to a smaller, fixed size for processing. A stem layer and multiple ccMamba layers are then used to encode this resized image data. The stem layer includes two consecutive convolutional operations, with the first convolutional operation using a stride of 2 to further reduce the image size.

Mamba leverages S4M to efficiently handle long sequence data, which can be represented as: 
\begin{gather}
x_{t}=Ax_{t-1} + Bx_{t} \notag \\
y_{t}=Cx_{t}   
\end{gather}
where A, B, and C are the parameters. $x_t$ and $y_t$ are the input and output temporal signals. 

To further reduce computational cost, we use a linear layer to compress the input channel dimension. Specifically, let $F^{G}$ and $F^{G'}$ represent the input and output feature maps with shape $(C, H, W, D)$. Given a token size $p$, the image feature map $F^{G}$ is converted into sequence data with shape $(L_G, T_G)$, where $L_G=\frac{HWD}{p^3}$ and $T_G=Cp^3$. We then compress the token channel to a fixed dimension $T$:
\begin{equation}
    F^{G} = \mathrm{LayerNorm}(\mathrm{Linear}(F^{G})) \in \mathbb{R}^{L_G \times T}
\end{equation}
Then two separate branches compute the intermediate results $F^{G}_1$ and $F^{G}_2$. They are combined through element-wise multiplication, which could be expressed as:
\begin{gather}
    F^{G}_1 = \sigma(\mathrm{Linear}(F^{G})) \notag \\ 
    F^{G}_2 = \mathrm{SSM}(\sigma(\mathrm{Conv}(\mathrm{Linear}(F^{G})))) \notag \\ 
    F^{G}_o = F^{G}_1 \odot F^{G}_2
\end{gather}
where $\sigma(\cdot)$ is the activation function $\mathrm{SiLU}$. Following the Transformer's architectural design, LayerNorm and an MLP with GELU activation are applied. A linear layer is finally used to recover the channel dimension.
\begin{equation}
    F^{G'} = \mathrm{Linear}(\mathrm{MLP}(\mathrm{LayerNorm}(F^{G}_o)))
\end{equation}

Let $F^{G}$ and $F^{L}$ represent the feature maps from the global branch and the decoder of the local branch, respectively. The goal of the CaM block is to enhance the representation capability of $F^{L}$ using $F^{G}$. The block consists of two main modules: Coordinate-Guided Local Feature Enhancement (Coord-LFE) and Coordinate-Aligned Global-Local Feature Fusion (Coord-GLF). These modules generate the features $F^{L|G}$ and $F^{L+G}$, respectively. The final output is produced by combining them with a convolution layer:
\begin{equation}
    F^{out} = \mathrm{Conv}([F^{L|G},F^{L+G}])
\end{equation}
where $F^{L|G}$ and $F^{L+G}$ are introduced in the following sections.
\subsubsection{Coordinate-Guided Local Feature Enhancement}
In this module, the features $F^{G}$ and $F^{L}$ are first tokenized. We then propose a unified positional encoding strategy to strengthen the spatial location relationship between the two feature maps. For each training iteration, we store the normalized center coordinates ($-1, 1$) of the randomly cropped patch relative to the full-resolution image.

Let $P_c: (x_o, y_o, z_o)$ denote the center coordinates of the patch. $(H, W, D)$ and $(h, w, d)$ represent the size of $F^{G}$ and $F^{L}$, respectively, while $p_G$ and $p_L$ are the token sizes for $F^{G}$ and $F^{L}$. For each token in $F^{L}$ with index $(i, j, k)$, the corresponding coordinate $P_t: (x, y, z)$ can be computed using the following formula:
\begin{gather}
x = x_o - \frac{h}{H} + (i + 0.5)\cdot p_L \cdot \frac{2}{H}  \notag \\
y = y_o - \frac{w}{W} + (j + 0.5)\cdot p_L \cdot \frac{2}{W}  \notag \\
z = z_o - \frac{d}{D} + (k + 0.5)\cdot p_L \cdot \frac{2}{D}
\end{gather}
The coordinates for each token in $F^{G}$ can also be calculated using this formula, with the center coordinate set to $(0, 0, 0)$. A linear layer is then applied to generate the positional embedding for each token. With this shared positional embedding, tokens with similar absolute coordinates in the global and local features will also have similar positional coordinates.

We then use a cross-attention mechanism to compute the coordinate-guided local feature $F^{L|G}$ based on the global feature. Specifically, we take the tokens of $F^{L}$ as queries to establish relationships between $F^{L}$ and $F^{G}$. This allows the local feature $F^{L}$ to perceive its global position, enabling the model to achieve a larger receptive field. This can be expressed as
\begin{gather}
F^{L|G} = \frac{Q^{L}K^{G}}{\sqrt{T}}V^{G}
\end{gather}
where $Q^{L}$ is obtained from $F^{L}$ through a linear layer, and $K^{G}$ and $V^{G}$ are derived from $F^{G}$ through two independent linear layers. $T$ represents the token dimension.
\subsubsection{Coordinate-Aligned Global-Local Feature Fusion}
Although $F^{L|G}$ captures relatively global information through coordinate guidance, it is derived from the coarse, low-resolution $F^{G}$, which may lack fine-grained local features. Therefore, we propose a coordinate-aligned feature fusion strategy to obtain the refined feature $F^{L+G}$. Inspired by \cite{hu2018squeeze, woo2018cbam, 10411857}, we first obtain the patch-local feature $F^{pG}$ by cropping $F^{G}$ at the center coordinate. The shape of $F^{pG}$ is kept the same as $F^{L}$. The initial fusion feature is computed simply as $F = F^{L} + F^{pG}$. This fusion feature is then used to compute the following attention masks for final feature fusion:

\textbf{Channel Attention}: The objective is to compute a one-dimensional channel-wise attention mask $M_c \in \mathbb{R}^{C}$ to re-calibrate the features. We first compress the spatial dimensions using a pooling layer, followed by two linear layers with $\mathrm{ReLU}$ activation to enhance expressive capability.
\begin{equation}
    M_c = W_2\cdot\mathrm{ReLU}(W_1\cdot \mathrm{Pooling}(F))
\end{equation}

\textbf{Spatial Attention}: The aim is to compute the spatial attention mask $M_s \in \mathbb{R}^{h\times w\times d}$. We first derive the spatial activation distribution across the channel dimension using channel-wise mean and max operations. A convolution layer with a large kernel size, followed by a sigmoid activation, is then applied to compute the spatial attention mask. This process is expressed as:
\begin{gather}
    F_{mean} = \mathrm{Mean}(F, 0), F_{max} = \mathrm{Max}(F, 0) \notag \\
    M_s = \mathrm{Sigmoid}(\mathrm{Conv}_{7}([F_{mean}, F_{max}]))
\end{gather}

\begin{table*}
\centering

\setlength{\tabcolsep}{2.5pt}
\begin{tabular}{lllllllllll}
\toprule
Datasets    & \multicolumn{5}{c}{PARSE}                                     & \multicolumn{5}{c}{KiPA}                                      \\
\cmidrule(lr){2-6} \cmidrule(lr){7-11}
Method      & Dice $\uparrow$ & NSD $\uparrow$ & ASSD $\downarrow$ & TRD $\uparrow$ & FPR $\downarrow$   & Dice $\uparrow$ & NSD $\uparrow$ & ASSD $\downarrow$ & TRD $\uparrow$ & FPR $\downarrow$        \\
\midrule
UNet        & 74.58±0.30 & 69.45±0.36 & 3.42±0.28 & 83.10±1.18 & 22.30±1.34 & 78.71±0.60 & 82.03±0.71 & 3.62±0.34 & \underline{90.37±0.68} & 21.34±0.97 \\
UNETR       & 69.87±1.30 & 63.88±0.64 & 3.67±0.28 & \textbf{87.67±0.66} & 33.01±2.57 & 78.51±1.45 & 81.77±1.01 & 3.71±0.21 & 90.35±0.55 & 22.19±1.71 \\
Swin UNETR   & 77.81±0.79 & 73.00±0.72 & 2.84±0.39 & \underline{84.66±1.04} & 19.18±1.61 & 81.30±0.42 & 84.41±0.72 & 3.12±0.32 & \textbf{90.63±1.29} & 18.31±0.71 \\
ERNet       & 79.32±0.54 & 74.61±0.89 & 2.13±0.19 & 72.39±5.86 & 10.27±2.52 & 79.98±1.30 & 84.23±1.87 & 2.76±0.61 & 88.62±2.26 & 16.90±4.94 \\
DSCNet      & 80.98±1.04 & 77.39±2.55 & 1.59±0.16 & 71.18±8.88 & \underline{7.71±3.67}\small{$^{\genfrac{}{}{0pt}{}{**}{\dagger\dagger}}$}  & 82.28±0.28 & 86.70±0.26 & 2.42±0.14 & 90.29±1.16 & 14.77±1.28 \\
TRA         & 73.24±0.66 & 66.44±0.30 & 2.86±0.14 & 77.05±1.70 & 22.13±0.54 & 78.36±0.91 & 82.12±0.50 & 3.10±0.24 & 83.35±3.88 & 17.49±4.07 \\
MDNet      & 76.10±1.58 & 70.06±1.73 & 2.40±0.29   & 73.28±9.22 & 15.23±1.56 & 81.69±0.33 & 85.46±0.31 & 2.63±0.23 & 89.33±1.63 & 16.18±0.50 \\
UMamba      & 82.13±0.85 & 81.32±1.16 & \underline{1.54±0.15}\small{$^{\dagger\dagger}$} & 79.81±2.25 & 9.96±1.95  & 82.04±0.50 & 87.07±2.21 & 2.37±0.57 & 87.69±0.30 & 13.39±2.00 \\
SegMamba    & 77.48±0.11 & 70.27±0.45 & 2.05±0.09 & 59.97±3.39 & 9.80±0.74  & \underline{83.18±1.14}\small{$^{\genfrac{}{}{0pt}{}{*}{\dagger}}$} & \underline{89.14±2.31}\small{$^{\genfrac{}{}{0pt}{}{**}{\dagger\dagger}}$} & \underline{1.81±0.94}\small{$^{\genfrac{}{}{0pt}{}{*}{\dagger\dagger}}$} & 83.73±0.90 & \underline{9.28±4.85}\small{$^{\genfrac{}{}{0pt}{}{*}{\dagger\dagger}}$}  \\
nnUNet      & \underline{82.87±0.07}\small{$^{\genfrac{}{}{0pt}{}{**}{\dagger\dagger}}$} & \underline{83.09±0.74}\small{$^{\genfrac{}{}{0pt}{}{**}{\dagger\dagger}}$} & 1.60±0.10 & 82.47±1.04 & 9.89±0.62  & 81.62±0.97 & 87.99±1.68 & 2.58±0.74 & 86.42±1.45 & 12.48±2.32 \\
COMMA       & \textbf{84.86±0.89} & \textbf{85.81±1.98} & \textbf{1.51±0.17} & 84.35±3.32 & \textbf{5.97±1.35}  & \textbf{84.88±1.30} & \textbf{91.04±1.58} & \textbf{1.05±0.33} & 85.07±2.17 & \textbf{6.31±1.39}  \\
\bottomrule
\end{tabular}
  \caption{The segmentation results on the PARSE (CT) and KiPA (CT) datasets are presented. The best and second-best results are highlighted in bold and underlined, respectively. Statistical significance: $*$, $**$ (pair-wise t-test) and $\dagger$, $\dagger \dagger$ (Wilcoxon signed-rank test) denote $p < 0.05$ and $p < 0.001$, respectively. p-values are adjusted for multiple hypotheses using Benjamini–Hochberg FDR (q=0.05).}
  \label{tab:comparison on CT}
\end{table*}

\begin{table*}
\centering

\setlength{\tabcolsep}{3pt}
\begin{tabular}{lllllllllll}
\toprule
Datasets    & \multicolumn{5}{c}{ASOCA}                                      & \multicolumn{5}{c}{TubeTK}                                       \\
\cmidrule(lr){2-6} \cmidrule(lr){7-11}
Method      & Dice $\uparrow$ & NSD $\uparrow$  & ASSD $\downarrow$ & TRD $\uparrow$ & FPR $\downarrow$     & Dice $\uparrow$ & NSD $\uparrow$  & ASSD $\downarrow$ & TRD $\uparrow$ & FPR $\downarrow$ \\
\midrule
UNet        & 76.62±2.71 & 79.45±2.36 & 4.13±0.47 & 90.87±1.76 & 20.68±1.96 & 72.24±0.11 & 87.07±0.30 & 0.91±0.12 & 84.09±0.49 & 8.51±0.60 \\
UNETR       & 75.15±1.59 & 78.44±0.83 & 4.51±0.20 & 91.03±1.06 & 23.26±1.01 & 74.51±0.62 & 87.65±0.42 & 0.80±0.02 & 84.97±0.40 & 8.13±0.54 \\
Swin UNETR   & 77.11±3.72 & 80.52±3.90 & 3.64±0.82 & 91.42±0.23 & 20.34±4.45 & 75.72±0.63 & 88.30±0.30 & 0.75±0.01 & 85.25±0.35 & 7.15±0.43 \\
ERNet       & 79.05±2.98 & 84.59±1.98 & 2.42±0.28 & 89.29±2.85 & 13.92±0.68 & 75.56±0.58 & 88.26±0.27 & 0.75±0.02 & 84.89±0.95 & 6.91±0.85 \\
DSCNet      & 80.16±2.69 & 86.10±0.96 & \underline{2.17±0.24} & 88.42±4.49 & \underline{11.74±1.96} & 75.19±0.01 & 87.82±0.07 & 0.77±0.00 & 84.56±0.19 & 6.91±0.01 \\
TRA         & 76.53±2.75 & 80.91±3.50 & 3.49±0.92 & 87.64±0.70 & 18.34±4.19 & 73.73±0.30 & 87.39±0.25 & 0.82±0.01 & 84.86±0.68 & 8.38±0.56 \\
MD-Net      & 77.98±1.73 & 83.24±2.22 & 2.82±0.87 & 86.21±5.42 & 14.79±5.32 & 75.53±0.19 & 87.84±0.21 & 0.77±0.00 & 83.01±0.37 & 6.19±0.01 \\
UMamba      & 80.29±2.38 & 85.30±1.98 & 2.33±0.35 & 90.05±2.45 & 13.60±0.57 & 74.21±0.33 & 87.42±0.19 & 0.80±0.02 & 83.68±0.95 & 7.35±0.61 \\
SegMamba    & 80.47±1.85 & 86.15±1.38 & 2.28±0.05 & 88.78±0.16 & 12.82±0.86 & 73.12±2.40 & 87.88±0.06 & 0.77±0.06 & 82.34±1.22 & \textbf{5.17±0.79} \\
nnUNet      & \underline{80.62±2.39}\small{$^{\genfrac{}{}{0pt}{}{**}{\dagger\dagger}}$} & \underline{86.58±1.74}\small{$^{\genfrac{}{}{0pt}{}{*}{\dagger}}$} & 2.68±0.43 & \underline{91.47±0.75} & 14.44±1.86 & \underline{77.30±0.76} & \underline{88.94±0.43}\small{$^{\genfrac{}{}{0pt}{}{*}{\dagger}}$} & \underline{0.72±0.01}\small{$^{\genfrac{}{}{0pt}{}{*}{\dagger\dagger}}$} & \textbf{85.70±0.11} & 6.42±0.37 \\
COMMA       & \textbf{82.27±1.68} & \textbf{87.86±0.71} & \textbf{1.95±0.26} & \textbf{91.57±0.32} & \textbf{11.58±0.52} & \textbf{78.04±0.32} & \textbf{89.38±0.03} & \textbf{0.70±0.01} & \underline{85.27±0.56} & \underline{5.28±0.56} \\
\bottomrule
\end{tabular}
  \caption{The segmentation results on the ASOCA (CT) and TubeTK (MRI) datasets are presented. Other symbols keep the same with \cref{tab:comparison on CT}.}
  \label{tab:comparison on CT-MRI}
\end{table*}

\begin{table*}
\centering

\setlength{\tabcolsep}{2.5pt}
\begin{tabular}{lllllllllll}
\toprule
Datasets    & \multicolumn{5}{c}{BreastMRI}                                      & \multicolumn{5}{c}{IXI}                                       \\
\cmidrule(lr){2-6} \cmidrule(lr){7-11}
Method      & Dice $\uparrow$ & NSD $\uparrow$  & ASSD $\downarrow$ & TRD $\uparrow$ & FPR $\downarrow$     & Dice $\uparrow$ & NSD $\uparrow$  & ASSD $\downarrow$ & TRD $\uparrow$ & FPR $\downarrow$ \\
\midrule
UNet        & 55.39±2.08 & 70.08±1.96 & 5.10±0.91 & 77.62±1.14 & 28.91±2.69 & 88.90±1.25 & 94.64±0.76 & 0.41±0.17 & 89.68±2.51 & 3.53±1.12 \\
UNETR       & 55.77±2.33 & 70.44±1.87 & 4.47±0.68 & 78.75±1.30 & 29.06±1.70 & 89.66±1.30 & 95.24±0.75 & 0.57±0.52 & 92.61±2.18 & 4.54±2.90 \\
Swin UNETR  & 58.36±2.44 & 72.75±2.20 & 4.26±0.67 & 78.93±3.26 & 25.75±1.22 & 91.29±1.22 & 96.17±0.46 & 0.23±0.02 & 92.96±2.08 & 2.45±0.31 \\
ERNet       & 57.95±3.60 & 71.77±3.95 & 4.82±0.59 & 75.43±4.73 & 25.95±3.17 & 90.85±1.26 & 96.21±0.44 & 0.23±0.02 & 92.70±1.59 & 2.26±0.13 \\
DSCNet      & 60.78±1.14 & 75.27±0.74 & 3.91±0.42 & \textbf{82.84±1.61} & 24.93±1.06 & 88.73±0.11 & 95.13±0.15 & 0.27±0.02 & 89.47±0.17 & 2.45±0.24 \\
TRA      & 55.44±2.07 & 69.89±1.71 & 4.75±0.48 & 77.24±1.05 & 29.83±2.06 & 88.43±1.55 & 94.25±0.83 & 0.32±0.03 & 87.52±2.70 & 2.65±0.30 \\
MDNet      & 58.04±3.73 & 72.40±4.21 & 4.14±0.25 & 72.87±7.86 & 19.96±1.52 & 90.90±1.50 & 96.03±0.52 & 0.23±0.02 & 91.43±1.93 & \underline{1.91±0.27}\small{$^{\genfrac{}{}{0pt}{}{**}{\dagger\dagger}}$} \\
UMamba      & 58.92±2.80 & 73.35±2.67 & 4.34±0.64 & 79.27±6.63 & 24.85±1.49 & 91.03±1.16 & 96.30±0.22 & 0.22±0.02 & 92.33±0.35 & 1.92±0.20 \\
SegMamba    & 56.15±0.62 & 71.84±1.68 & 3.97±0.82 & 68.07±3.19 & \textbf{16.74±1.37} & 91.19±1.09 & 96.31±0.39 & 0.23±0.04 & 92.68±1.41 & 2.18±0.32 \\
nnUNet      & \underline{61.27±2.00}\small{$^{\genfrac{}{}{0pt}{}{**}{\dagger\dagger}}$} & \underline{76.30±2.41}\small{$^{\genfrac{}{}{0pt}{}{**}{\dagger\dagger}}$} & \underline{3.82±0.77}\small{$^{*}$} & \underline{82.21±1.16} & 21.77±2.33 & \underline{92.03±1.09}\small{$^{\genfrac{}{}{0pt}{}{**}{\dagger\dagger}}$} & \underline{96.76±0.43}\small{$^{\genfrac{}{}{0pt}{}{**}{\dagger\dagger}}$} & \underline{0.20±0.02}\small{$^{\genfrac{}{}{0pt}{}{*}{\dagger\dagger}}$} & \underline{93.92±1.81}\small{$^{\dagger}$} & 2.00±0.29 \\
COMMA       & \textbf{63.72±0.49} & \textbf{78.27±0.53} & \textbf{3.11±0.61} & 80.17±4.09 & \underline{18.31±3.35} & \textbf{92.11±1.64} & \textbf{96.90±0.52} & \textbf{0.18±0.03} & \textbf{94.11±1.75} & \textbf{1.81±0.12} \\
\bottomrule
\end{tabular}
  \caption{The segmentation results on the BreastMRI (MRI) and IXI (MRI) datasets are presented. Other symbols keep the same with \cref{tab:comparison on CT}.}
  \label{tab:comparison on MRI}
\end{table*}

\begin{table}
\centering

\begin{tabular}{llll}
\toprule
Method      & FLOPs/G           & GPU usage/M           & Param./MB \\
\midrule
UNet        & \underline{204.32}      & \textbf{1581.47}      & 7.91      \\
UNETR       & 281.01            & 4875.40               & 169.03    \\
SwinUNETR  & 329.34            & 9134.42               & 61.99     \\
ERNet       & \textbf{171.14}   & 3258.77               & \textbf{5.11}      \\
DSCNet      & 221.97  & 48124.98                    & \underline{7.20}      \\
TRA         & 320.45  & \underline{2056.81}                     & 45.60     \\
MDNet      & 2220.05 & 8986.06                 & 157.86    \\
UMamba      & 3814.20 & 12954.88                & 173.53    \\
SegMamba    & 655.49  & 6552.42                 & 64.24     \\
nnUNet      & 1529.51 & 5725.46                 & 30.80     \\
COMMA       & 249.31  & 8833.45                 & 44.96    \\
\bottomrule
\end{tabular}
  \caption{Computational cost metrics for various methods. Other symbols keep the same with \cref{tab:comparison on CT}.}
  \label{tab:comparison on complexity}
\end{table}

\textbf{Pixel Attention}: The pixel-level attention mask is computed using $M_c$, $M_s$, and $F$. We first combine the channel and spatial masks by simple addition. To obtain the final mask, the combined mask and features are concatenated and processed through a convolutional operation. This can be expressed as:
\begin{equation}
    M_p = \mathrm{Sigmoid}(\mathrm{Conv}_{7}([M_c + M_s, F]))
\end{equation}
The fusion feature is then computed with $M_p$:
\begin{equation}
    F^{L+G} = \mathrm{Conv}(M_p\cdot F^{pG} + (1-M_p)\cdot F^{L}) + F^{L}
\end{equation}
\subsection{Loss Function}
The local and global branches generate the patch and global predictions using a convolutional operation. The segmentation loss for the local branch can be expressed as:
\begin{equation}
    \mathcal{L}_{L} = \mathrm{Diceloss}(\hat{Y}_L, Y_L) + \mathrm{BCEloss}(\hat{Y}_L, Y_L)
\end{equation}
where $\hat{Y}_L$ and $Y_L$ represent the patch prediction and patch segmentation mask, respectively. Similarly, the segmentation loss for the global branch, denoted as $\mathcal{L}_{G}$, is computed by applying the same formula to the global branch's output and the downsampled global ground truth. The final loss function is a linear combination of $\mathcal{L}_{L}$ and $\mathcal{L}_{G}$:
\begin{equation}
    \mathcal{L} = \mathcal{L}_{L} + \lambda\mathcal{L}_{G}
\end{equation}

\section{Experimental Setup}
\subsection{Datasets}
Six vessel segmentation datasets are used for evaluation in this study, including three CT datasets and three MRI datasets:
\begin{itemize}
    \item \textbf{KiPA}: The dataset is sourced from "the Kidney Parsing Challenge 2022" \cite{HE2021102055}. The goal is to segment renal vein structures from CT images. Renal vein structures contain only a small number of tracers, usually exhibiting low contrast. A total of 70 cases are used in this study. The image sizes range from $(116, 116, 155)$ to $(179, 179, 280)$. 56 cases are used for training, and 14 cases are used for testing.
    \item \textbf{ASOCA}: This dataset originates from the "Automated Segmentation Of Coronary Arteries" challenge \cite{gharleghi2022computed}. The objective is to segment coronary arteries from cardiac computed tomography angiography images. The dataset includes 40 cases, with 20 from healthy patients and 20 from patients diagnosed with coronary artery disease. The Image sizes range from $(512, 512, 168)$ to $(512, 512, 224)$. 32 cases are used for training, and 8 cases are used for testing.
    \item \textbf{PARSE}: This dataset is from the "Pulmonary Artery Segmentation Challenge 2022", targeting the segmentation of pulmonary artery structures from CT images with high accuracy \cite{luo2023efficient}. A total of 100 publicly accessible cases are used in this study. The image sizes range from $(512, 512, 228)$ to $(512, 512, 390)$. 80 cases are used for training, and 20 cases are used for testing.
    \item \textbf{TubeTK}: This dataset includes 42 TOF-MRA images from healthy patients \footnote{\url{https://public.kitware.com/Wiki/TubeTK/Data}}, with the goal of segmenting cerebrovascular structures. The annotation mask is sourced from \cite{shi2023benefit}. All images in this dataset have a consistent size of $(448, 448, 128)$. 33 cases are used for training, and 9 cases are used for testing.
    \item \textbf{BreastMRI}: This dataset comes from \cite{lew2024publicly} with 100 cases. The objective is to segment blood vessels within the breast, with images acquired using a 1.5 T or 3.0 T scanner. Three cases are excluded from this study due to mismatched shapes between the image and the mask. The Image sizes range from $(320, 320, 144)$ to $(512, 512, 200)$. 80 cases are used for training, and 17 cases are used for testing.
    \item \textbf{IXI}: The image data are sourced from the IXI dataset \footnote{\url{http://brain-development.org/}}, aiming to segment cerebrovascular structures from MRA images. This dataset comprises 570 cases, with manually annotated masks provided in this study \footnote{https://github.com/shigen-StoneRoot/COMMA}. The majority of images (497 cases) have a size of $(512, 512, 100)$, while 27 cases have a size of $(1024, 1024, 92)$, and one case is sized $(1024, 1024, 91)$. 
\end{itemize}
\begin{figure*}
  \centering
    \centering
    \includegraphics[scale=0.25]{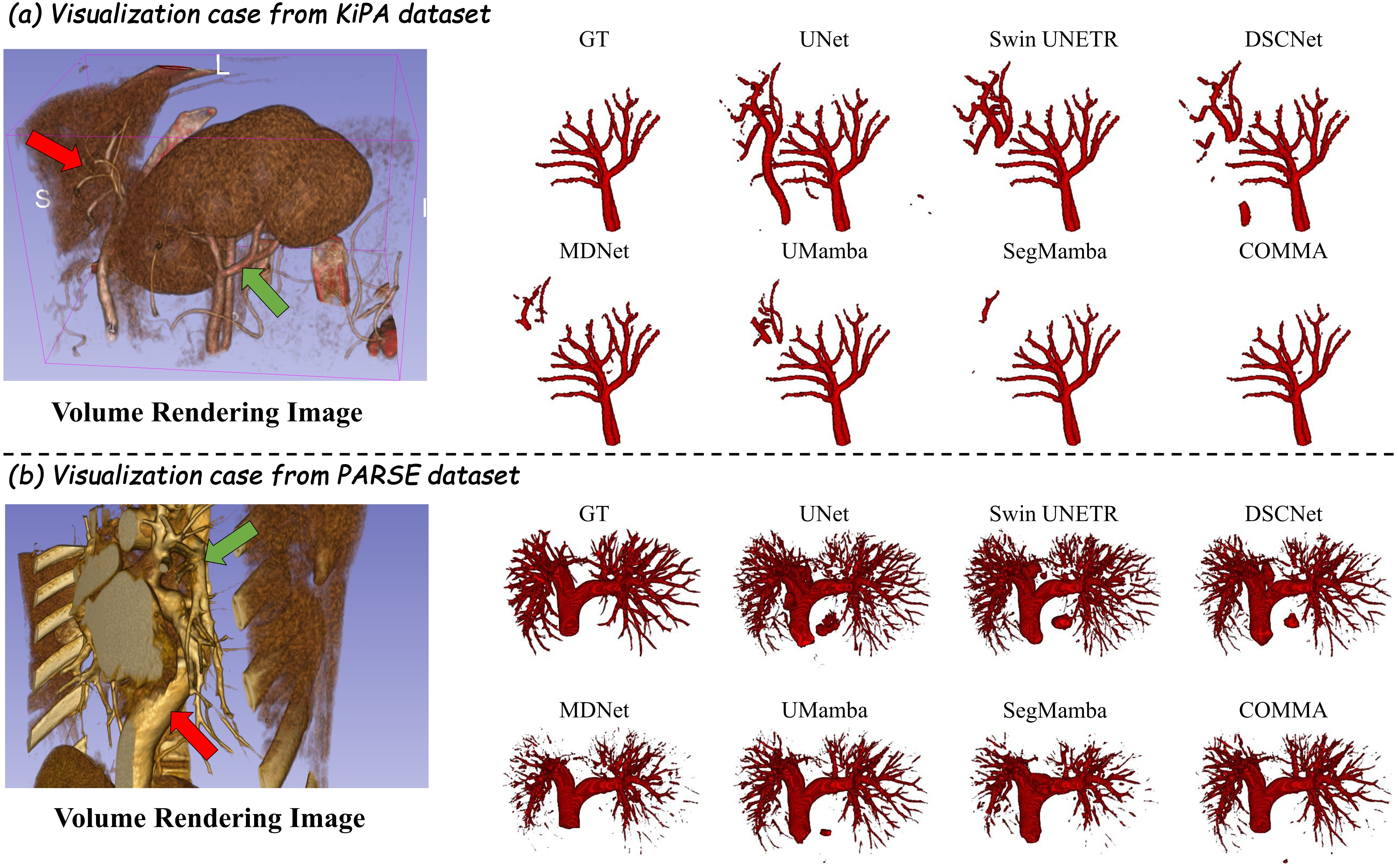}
    \caption{The 3D visualization results from (a) KiPA and (b) PARSE datasets. The green arrows indicate the segmented target vessels, while the red arrows highlight the vascular structures outside the target region, representing the artifacts produced by the segmentation.}
    \label{fig: visu_CT}
\end{figure*}
\subsection{Evaluation Metrics}
In this study, the primary evaluation metrics are: Dice similarity coefficient (Dice), average symmetric surface distance (assd), Normalized Surface Dice (NSD), tree length detected (TRD), and false positive rate (FPR). These metrics are selected for their ability to comprehensively evaluate segmentation accuracy, centerline precision, and vessel connectivity. Due to space constraints, additional metrics, including branch count, branches detected, tree length, leakage count, leakage volume \cite{lo2012extraction}, centerline Dice coefficient (clDice) \cite{cldice2021}, and HD95 are reported in the supplementary material.

Additionally, we compute the FLOPs (floating point operations), model parameters, and the maximum GPU memory usage during training to assess the computational complexity of the models. 
\subsection{Competing Models}

We evaluate our method against a range of baseline models, including CNN-based architectures, Transformer-based designs, and models specifically tailored for vessel segmentation. Additionally, we include recent Mamba-based segmentation models to ensure a comprehensive comparison. The competing methods are summarized as follows:

\begin{itemize}
    \item \textbf{UNet} \cite{ronneberger2015u}: UNet is the classical encoder–decoder architecture widely used in medical image segmentation tasks. The architecture features symmetric skip connections between encoder and decoder layers, which help preserve spatial resolution and facilitate gradient flow. 
    
    \item \textbf{UNETR} \cite{hatamizadeh2022unetr}: UNETR employs a transformer-based encoder to model global and multi-scale features from volumetric inputs. It connects the transformer with a convolutional decoder through multi-resolution skip connections, following a U-Net-like architecture for semantic segmentation.
    
    \item \textbf{SwinUNETR} \cite{tang2022self}: Swin UNETR extends UNETR by incorporating Swin Transformer blocks, which introduce hierarchical and shifted window attention. This allows the model to better capture local structures and convolutional priors, achieving a good trade-off between performance and computational efficiency.
    
    \item \textbf{ERNet} \cite{xia20223d}: ERNet incorporates a reverse edge attention module, edge-enhanced loss, and feature selection to better capture vascular boundaries and address the imbalance between edge and non-edge voxels.
    
    \item \textbf{DSCNet} \cite{qi2023dynamic}: DSCNet leverages the continuity and elongation of tubular structures through all stages of the network. It incorporates dynamic snake convolution, a multi-view feature integration approach, and a topology-aware loss function to improve segmentation accuracy and preserve structural consistency in both 2D and 3D scenarios.
    
    \item \textbf{TRA} \cite{chen2023cerebrovascular}: A topology-constrained generative framework developed specifically for segmenting cerebral vessels in TOF-MRA volumetric data. It leverages self-supervised spatial layout learning and incorporates vascular skeleton features to preserve topological continuity.
    
    \item \textbf{MDNet} \cite{huang2025representing}: MDNet incorporates fractal feature maps to capture the self-similar topology of tubular structures. It enhances U-Net with dedicated edge and skeleton decoders, and integrates fractal dimension–based features into both the input and loss function to improve boundary precision and vascular continuity.
    
    \item \textbf{UMamba} \cite{ma2024u}:  A general-purpose biomedical segmentation model that integrates convolutional layers with state space models to jointly capture local features and long-range dependencies, offering improved efficiency and performance over traditional CNNs and Transformers.
    
    \item \textbf{SegMamba} \cite{xing2024segmamba}: A 3D medical image segmentation model built upon state space models, designed to replace Transformer-based modules for more efficient long-range dependency modeling.

    \item \textbf{nnUNet} \cite{isensee2021nnu}: nnUNet is an adaptive and self-configuring deep learning framework designed for medical image segmentation. It automatically configures the network architecture, preprocessing pipeline, and postprocessing steps based on the dataset at hand, allowing it to perform well on a wide variety of segmentation tasks.
\end{itemize}

\subsection{Implementation Details}
For the IXI dataset, which includes 570 cases, we allocate 400 cases for training, 56 for validation, and 114 for testing. The model is trained using the SGD optimizer for 25,000 iterations with a batch size of 2. For datasets with smaller sample sizes, 80\% of the total cases are used for training, with the remaining cases reserved for testing, based on the model’s final iteration. To ensure robustness, the dataset is randomly split three times, and we report the average results across these three splits. The local branch is based on randomly cropped patches with a size of (96, 96, 96). The encoder consists of four stages, with initial channel dimensions of 32, followed by [64, 128, 256, 320]. In the global branch, the downsampled image size is (256, 256, 96). The stem layer includes two convolutional layers with channel dimensions of 16 and 32, respectively, and a stride of 2 in the XY plane for the first convolution. The ccMamba block has a fixed channel dimension (T) of 32. For the CaM block, the local feature patch sizes for each stage are [1, 2, 3, 6], while the global feature patch size remains consistently at 8. The number of attention heads is set to 8. The value of $\lambda$ is empirically set to 0.25. Regarding the inference protocol, both the local and global branches contribute to the prediction during the testing phase. Specifically, local-patch logits are overlapped and blended using a Gaussian blending technique, with a stride of 0.5 times the patch size. The Gaussian blending technique is applied to the logits to ensure smooth integration. 

Our experiments were primarily conducted on an AMD Ryzen Threadripper PRO 5955WX processor with an RTX 4090 GPU for both training and testing. This configuration was used for nearly all of our experiments. For DSCNet baseline, due to memory limitations, we conducted the experiments on an alternate setup consisting of an AMD EPYC 9174F 16-core processor paired with an NVIDIA A100 GPU. This adjustment was necessary to accommodate DSCNet’s larger memory requirements during training. Additionally, we used mixed-precision training to optimize memory usage and computational efficiency during model training.
\begin{figure*}
\centering
\includegraphics[scale=0.6]{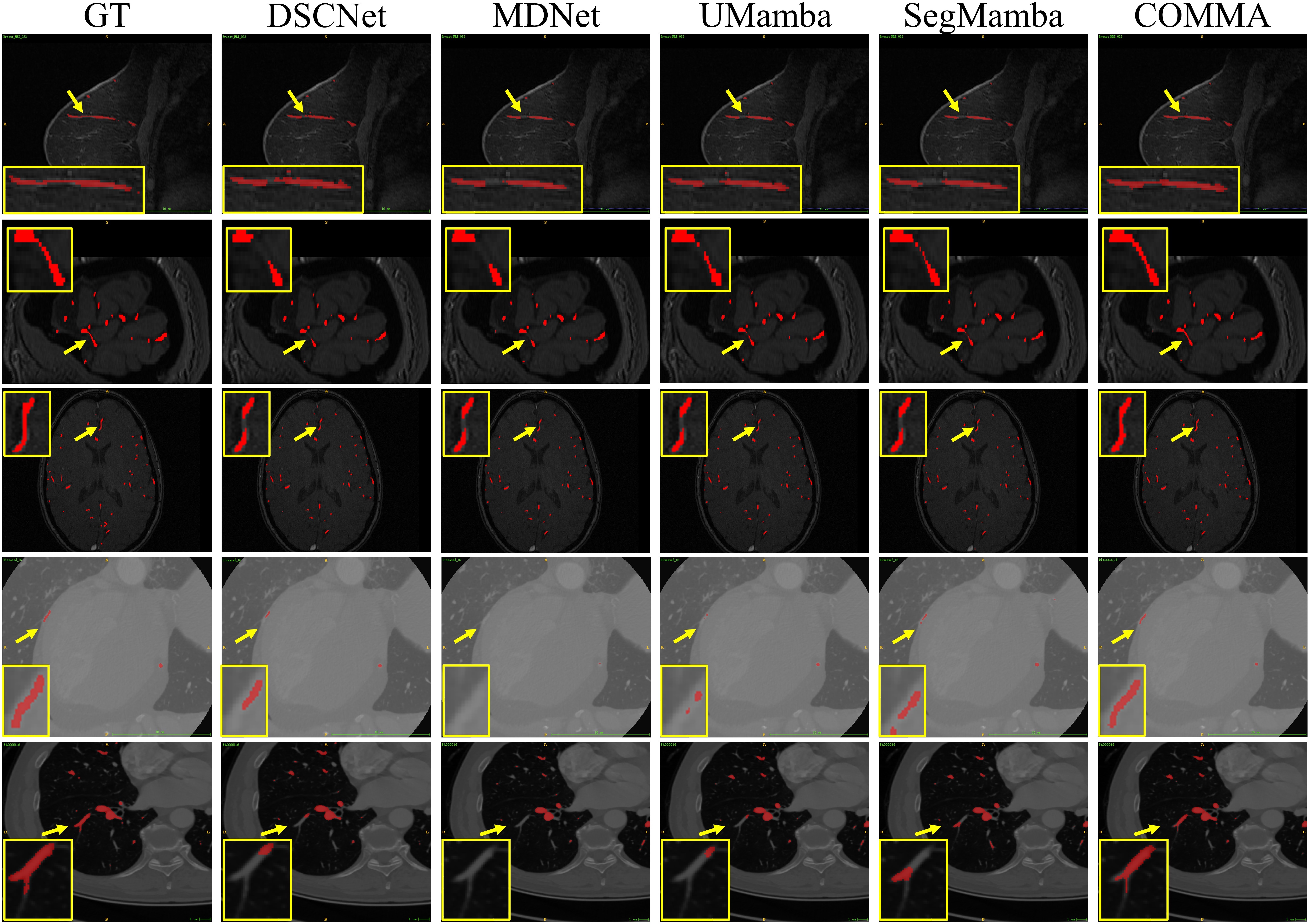}
\caption{The slice visualization results to illustrate the intuitive evaluation. The datasets from top to bottom are BreastMRI, TubeTK, IXI, ASOCA, and PARSE.}
\label{fig: visu_MRI}
\end{figure*}
\subsection{Definition of Sparsity Index and Dispersion Index}
\label{sec definition Dispersion Index}
Let $\mathcal{S}={(I_n, Y_n)}$ represent a segmentation dataset, where $I_n$ and $Y_n$ denote the image and mask, respectively. The Sparsity Index and Dispersion Index are defined in relation to $Y_i$ and can be computed as follows:
\begin{equation}
    SI = \frac{\sum_{i=1}^H\sum_{j=1}^W\sum_{k=1}^D Y_n(i, j, k)}{H\times W\times D}
\end{equation}
where $(H, W, D)$ represents the dimensions of $Y_i$. The Sparsity Index quantifies the ratio of non-zero elements within the segmentation mask, without regard to their spatial distribution. The Dispersion Index, on the other hand, characterizes the spatial spread of these non-zero elements.

Let $M_n = { (i, j, k) \mid Y_n(i, j, k) \neq 0 }$ denote the set of coordinates of non-zero elements. First, we calculate the center point coordinate:
\begin{equation}
    (i_c, j_c, k_c) = \frac{\sum\limits_{(i, j, k) \in M_n} (i, j, k)}{\lvert M_n \rvert}
\end{equation}
Next, we compute the distance from each non-zero element to this center point:
\begin{equation}
    D(i, j, k) = \sqrt{(i-i_c)^2 + (j-j_c)^2 + (k-k_c)^2} 
\end{equation}
Finally, the Dispersion Index is determined by the normalized mean distance:
\begin{equation}
    DI = \frac{\sum\limits_{(i, j, k) \in M_n} 2 \cdot D(i, j, k)}{\sqrt{H^2 + W^2 + D^2}}
\end{equation}
The $DI$ takes into account the spatial arrangement of non-zero elements. For compact organs, where non-zero elements are close together, the normalized mean distance—and consequently $DI$—will be low. In contrast, dispersed organs exhibit a larger $DI$.
\section{Results}
\subsection{Comparison with Competing Methods}
The vessel segmentation results for all datasets (PARSE, KiPA, ASOCA, TubeTK, BreastMRI, and IXI) are presented in \cref{tab:comparison on CT}, \cref{tab:comparison on CT-MRI}, and \cref{tab:comparison on MRI}. Statistical significance was assessed using paired t-tests and Wilcoxon signed-rank tests for paired comparisons, conducted over three randomly split test sets. The sample sizes for each dataset were as follows: KiPA (42), ASOCA (24), PARSE (60), TubeTK (24), BreastMRI (58), and IXI (342). On all six datasets, COMMA outperforms the baseline methods across most evaluation metrics, with significant statistical differences in many cases. For example, on the PARSE dataset, COMMA achieves the highest Dice and NSD scores, surpassing all other methods, with improvements of up to 1.5\% in Dice and 1.3\% in NSD compared to the second-best methods. On the IXI dataset, which includes high-quality annotations and a large number of training samples, COMMA also shows a modest advantage.

It is worth noting that KiPA exhibits a slightly different pattern in structure-level metrics: although COMMA achieves strong overlap/surface metrics and lower false-positive behavior, its TRD is not always the highest. A key reason is the trade-off between branch-detection sensitivity and false-positive control. For KiPA dataset, we only segment the renal vein, which is characterized by low-significance venous contrast, strong background interference, and substantial shape variation caused by kidney/tumor conditions. Consequently, more aggressive branch extension can increase detected tree length while also introducing non-target vascular-like regions, thereby increasing leakage and FPR. By contrast, COMMA is relatively conservative in uncertain regions due to coordinated global–local constraints, which improves specificity and suppresses out-of-target predictions, but may miss some marginal connectivity that contributes to TRD.

\cref{tab:comparison on complexity} illustrates the computational complexity of the models. COMMA also maintains computational efficiency even while simultaneously utilizes the patch data and entire image, compared with top-performing baselines. Specifically, COMMA’s FLOPs are one order of magnitude lower than those of MDNet and UMamba, and are half of SegMamba's. Additionally, COMMA’s peak GPU usage is lower than that of UMamba and SegMamba, with maximum GPU memory consumption only one-sixth of DSCNet’s.
\begin{table}
\centering
\begin{tabular}{lllllll}
\toprule
Dataset     & \multicolumn{3}{c}{PARSE(Branch Vessel)}         & \multicolumn{3}{c}{IXI (Distal Vessel)}          \\  \cmidrule(lr){2-4} \cmidrule(lr){5-7}
Method      & Dice           & clDice         & NSD            & Dice           & clDice         & NSD            \\ \midrule
UNet        & 64.55          & 62.62          & 71.88          & 83.60          & 84.15          & 92.07          \\
UNETR       & 57.53          & 58.52          & 65.25          & 86.00          & 87.38          & 94.14          \\
Swin UNETR  & 64.92          & 63.26          & 71.93          & 86.84          & 88.54          & 94.49          \\
ERNet       & 67.83          & 66.23          & 75.71          & 86.24          & 88.69          & 94.73          \\
DSCNet      & \underline{73.37}    & \underline{71.62}    & \underline{80.42}    & 85.62          & 88.07          & 94.21          \\
TRA  & 53.92          & 55.66          & 65.38          & 82.73          & 84.17          & 91.93          \\
MDNet      & 55.71          & 55.85          & 66.52          & 85.92          & 88.75          & 94.26          \\
UMamba      & 72.27          & 70.31          & 80.27          & 86.31    & \underline{89.63}    & \underline{94.92}    \\
SegMamba    & 59.58          & 56.67          & 67.27          & \underline{86.75}          & 88.98          & 94.85          \\
COMMA       & \textbf{76.23} & \textbf{75.75} & \textbf{84.32} & \textbf{88.07} & \textbf{90.05} & \textbf{95.26} \\ \bottomrule
\end{tabular}
  \caption{The segmentation results of the small vessels on the PARSE and IXI datasets.}
  \label{tab: small vessel}
\end{table}
\begin{figure}
  \centering
    \centering
    \includegraphics[width=\linewidth]{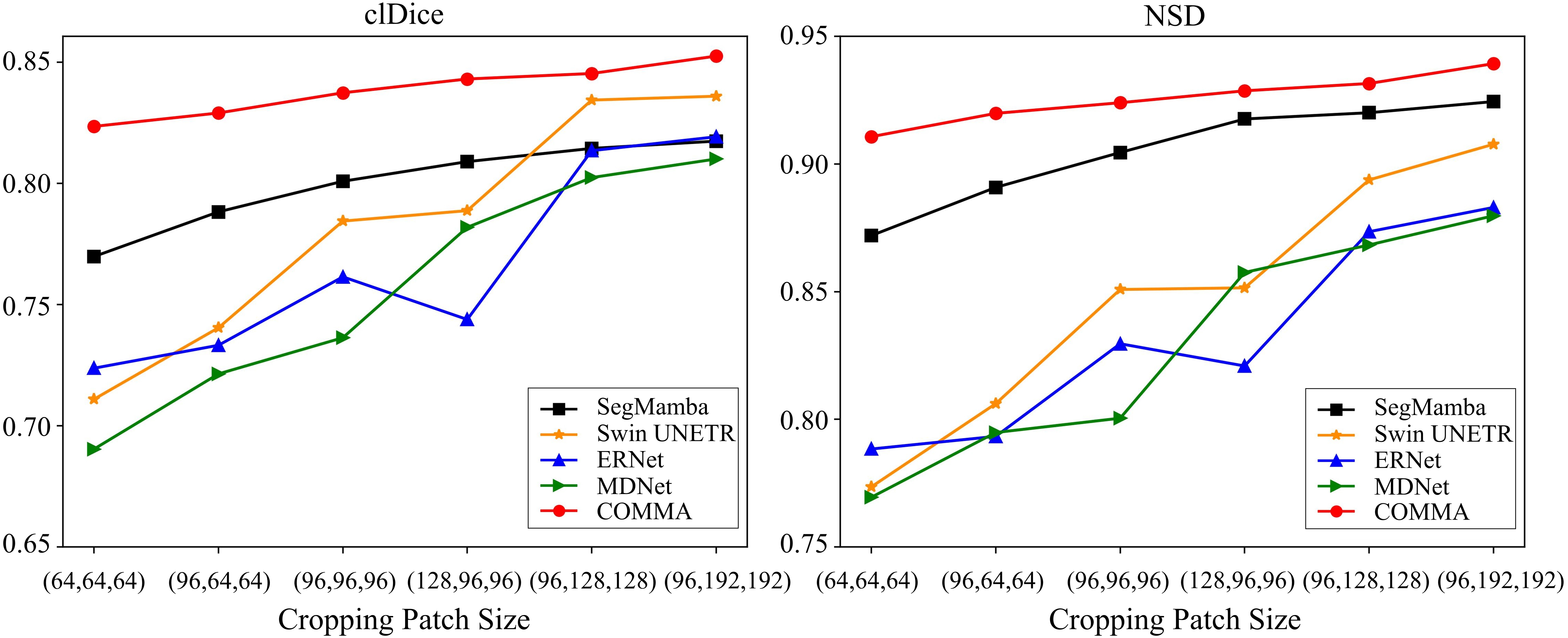}
    \caption{Effect of patch size variation across different methods on the KiPA dataset.}
    \label{fig: patch_size}
\end{figure}
\subsection{Visualization}
We present 3D visualization results for an intuitive evaluation in \cref{fig: visu_CT}. In the KiPA dataset example, baseline models generate varying levels of artifacts, while COMMA produces significantly fewer artifacts. These artifacts, appearing as vascular-like structures outside the target region, are highlighted by the red arrows, with the green arrows indicating the intended target regions. The artifacts are likely due to the absence of spatial relationship modeling in the baseline methods, which increases the likelihood of erroneously segmenting vessels outside the target area. This observation underscores the advantage of our method, which effectively minimizes such false-positive regions by leveraging spatial context and better capturing the anatomical structures within the target region. A similar trend is observed in the PARSE dataset, where segmentation masks from MDNet and SegMamba contain fewer artifacts but lack the level of vessel detail achieved by COMMA. Slice-view visualizations are provided in \cref{fig: visu_MRI}, where COMMA demonstrates superior performance in preserving vessel structure continuity. COMMA consistently achieves more continuous vessel structures compared to the top-performing baselines.

\subsection{Results on Small Vessel Segmentation}
Small vessels are typically more uncertain in spatial location compared to larger vessels (e.g., the Circle of Willis in cerebrovascular structures). We examine the effectiveness of COMMA in segmenting small vessel structures. In this study, we define the vessels inside the lung as small vessels (also called branch vessels). This definition follows the study \cite{luo2023efficient}. Besides, the distal vessels in the cerebrovascular (those with z-axis $>$ 60) are also considered small vessels for IXI dataset. The results are shown in \cref{tab: small vessel}.

COMMA achieves approximately 1.66\% higher performance than DSCNet for full vessel segmentation in terms of Dice, with the gap increasing to 2.86\% for small vessel segmentation. Similarly, in the IXI dataset, the performance difference between COMMA and SegMamba is 0.84\% for full vessel segmentation, widening to 1.76\% for small vessel segmentation.

\subsection{Impact of Patch Size on Spatial Awareness}
Patch size plays a crucial role in determining the extent of spatial context available to segmentation models. Smaller patches inherently contain less global information and are more prone to spatial ambiguity, especially when segmenting dispersed vascular structures. In this section, we evaluate the robustness of COMMA under varying patch sizes to assess how well it maintains spatial awareness.

As shown in \cref{fig: patch_size}, COMMA consistently outperforms baseline methods across all patch sizes. Notably, as the patch size decreases (e.g., $64^3$), COMMA exhibits a significant performance advantage in terms of both Dice and clDice scores. This robustness can be attributed to the model's ability to encode the relative spatial positions of each patch within the global image. In contrast, traditional patch-wise models such as SegMamba and Swin UNETR suffer a noticeable performance drop with smaller patches, as they lack access to global spatial cues. These results highlight the importance of incorporating spatial location awareness in patch-based processing. Even at a relatively larger patch size of $(96, 192, 192)$, our method maintains its superior performance, further demonstrating the effectiveness of our dual-branch strategy in preserving spatial context.

\begin{table}
\centering
\tabcolsep=1mm
\begin{tabular}{lllllll}
\toprule
Method                          & Coord-LFE & Coord-GLF & GLB Loss & Dice  & clDice & NSD   \\ \midrule
\multirow{4}{*}{Variants}   & \ding{56}         & \ding{56}         & \ding{56}           & 80.88 & 70.62  & 78.65 \\ 
 & \ding{56}         & \ding{52}         & \ding{52}           & 82.35 & 74.92  & 81.68 \\
                                & \ding{52}         & \ding{56}         & \ding{52}           & 82.34 & 73.87  & 81.24 \\
                                & \ding{52}         & \ding{52}         & \ding{56}           & 82.55 & 75.55  & 82.13 \\ \hline
COMMA                                & \ding{52}         & \ding{52}         & \ding{52}           & \underline{83.84} & \underline{76.96}  & \underline{83.54} \\ 
Rand. Coord.      & \ding{52}         & \ding{52}         & \ding{52}           & 81.45 & 75.31  & 81.35 \\ 
Physical Coord.      & \ding{52}         & \ding{52}         & \ding{52}           & \textbf{83.98} & \textbf{77.55}  & \textbf{83.58} \\
\bottomrule
\end{tabular}
  \caption{Ablation study results on the PARSE dataset. 'Rand. Coord.' refers to the COMMA model with randomized coordinates instead of precise calculations in \cref{sec: The Global Branch}, during the training phase. 'Physical Coord.' refers to the COMMA model utilizing real-world physical coordinates rather than index-based coordinates.}
  \label{tab: ablation study}
\end{table}
\begin{table}
\centering
\begin{tabular}{lllllll}
\toprule
GLB Branch  & FLOPs/G & GPU/G & Dice  & clDice & NSD   \\ \midrule
Mamba       & \textbf{249.31}   & \textbf{8.63}        & \textbf{86.36} & \textbf{84.31}  & \textbf{92.87} \\
Transformer & 366.61   & 62.23       & 85.50 & 83.16  & 90.11 \\ \bottomrule
\end{tabular}
  \caption{Comparison of GPU usage between Transformer and Mamba in the global branch on the KiPA dataset.}
  \label{tab: computation cost}
\end{table}
\subsection{Ablation Study}
\subsubsection{Contribution of Different Model Components}
\begin{figure}
  \centering
  \includegraphics[width=\linewidth]{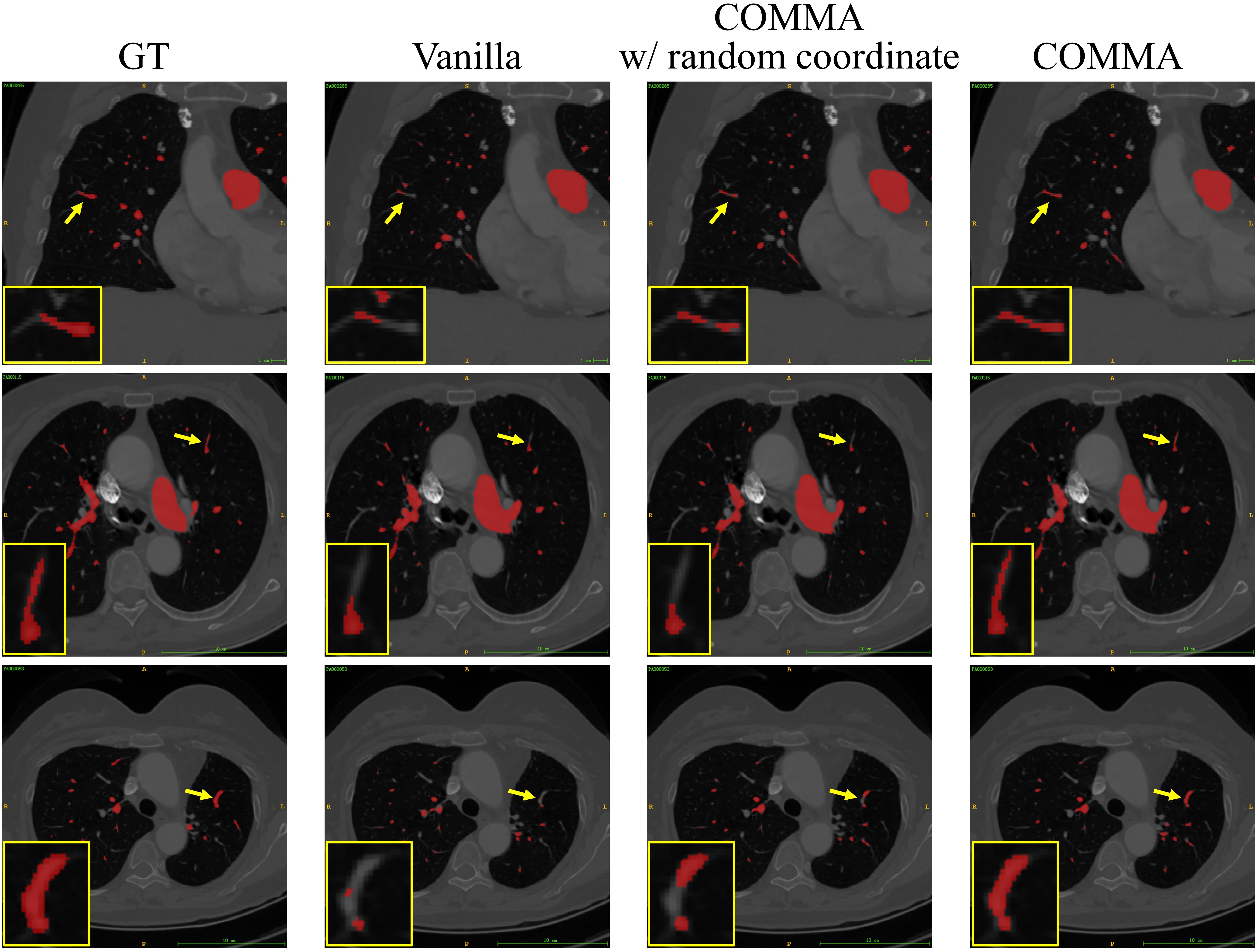}
    \caption{Slice visualization results for different COMMA variants. The yellow insets highlight enhanced continuity in local vessel structures.}
    \label{fig: ablation visu}
\end{figure}

We conduct experiments to evaluate the effectiveness of our proposed modules in \cref{tab: ablation study}. The vanilla model without any modules shows lower Dice, clDice, and NSD scores. Adding Coord-GLF and Global Loss significantly improves performance, with increases of 2-3\% in each metric. Similarly, incorporating Coord-LEF and Global Loss also enhances model performance. Including all components further boosts performance, with increases of 3-5\% across the metrics. Besides, omitting Global Loss leads to decreased model performance, underscoring its importance. Notably, when the coordinates are randomized instead of precise calculations in \cref{sec: The Global Branch}, model performance decreases by approximately 2\%, highlighting the importance of spatial location information. Furthermore, 'Physical Coord.' refers to the use of real physical coordinates derived from the 3D image's origin and spacing information, as opposed to index-based coordinates. The results show no significant improvement in the performance metrics, which aligns with our expectations. This is because our method primarily leverages relative positional relationships, and both real physical coordinates and index-based coordinates can effectively construct these relative relationships. The visualization results of COMMA variants in slice view is shown in \cref{fig: ablation visu}. Compared with its variant models, COMMA with coordinate guidance shows an advantage in segmenting small vessels farther from the center. 
\subsubsection{Computational Efficiency: Mamba vs. Transformer}
We also conduct an experiment to compare the computational efficiency of the Mamba structure with the Transformer structure. As shown in \cref{tab: computation cost}, both structures achieve comparable segmentation performance; however, the Transformer structure consumes significantly more GPU memory than Mamba. 

\subsubsection{Impact of Stride Size on Model Performance}
\begin{figure}
  \centering
  \includegraphics[scale=0.185]{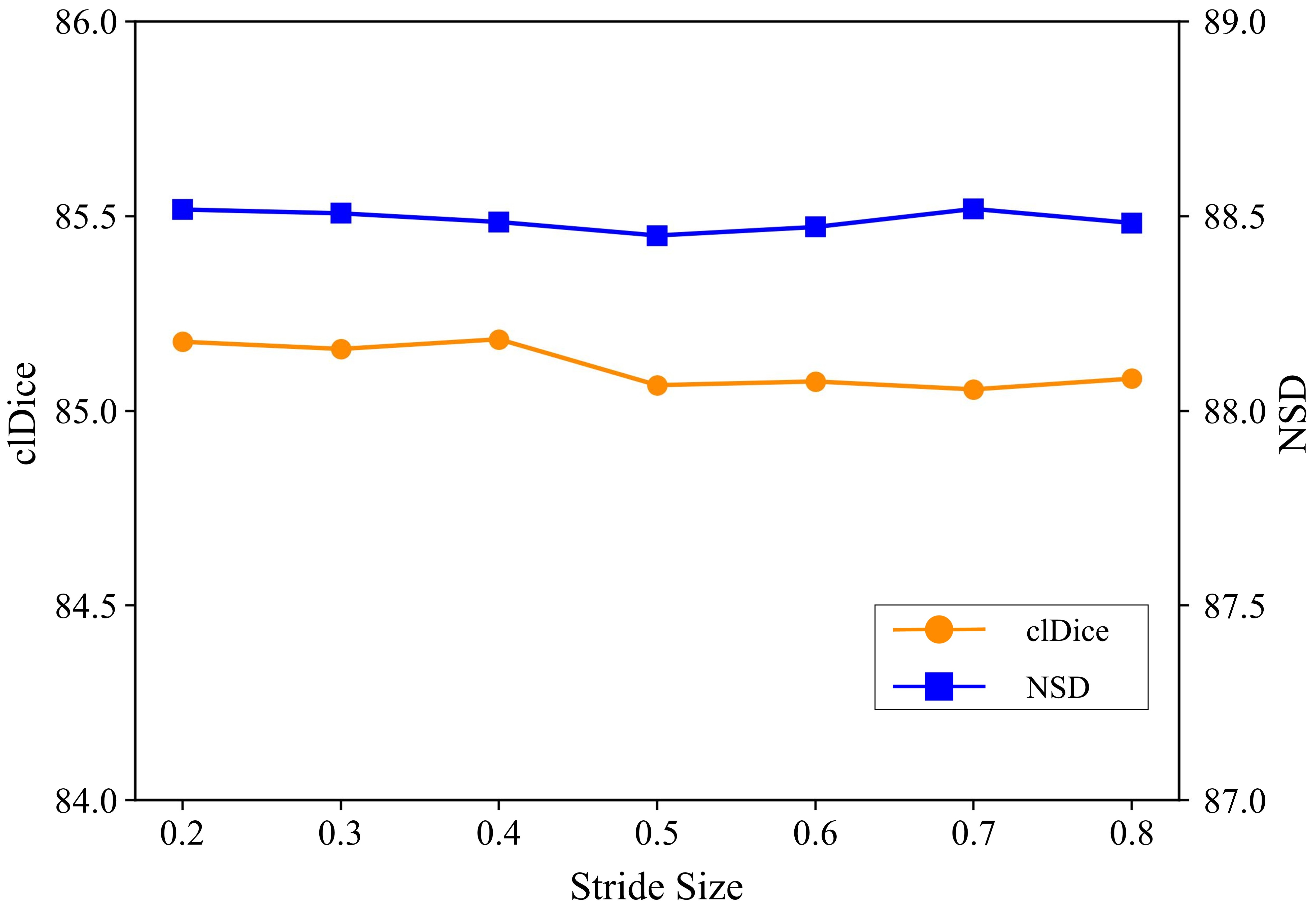}
    \caption{Effect of stride size during inference on model performance on the TubeTK dataset.}
    \label{fig: stride_size}
\end{figure}

In this experiment, we investigate the effect of varying stride sizes during inference on the model's performance, as shown in \cref{fig: stride_size}. Specifically, we evaluate how different stride sizes, ranging from 0.2 to 0.8, influence the clDice and NSD metrics. The results, obtained from the TubeTK dataset, demonstrate that stride size has a minimal impact on both metrics, with performance remaining relatively consistent across all tested stride values. This suggests that the model's ability to capture relevant features is not significantly affected by stride size, indicating a robust performance despite varying stride configurations.

\subsubsection{Complexity analysis of CaM}
In this section, we evaluate the computational complexity and performance trade-offs of the CaM module by examining the impact of patch size and token size on computational cost and model accuracy. The first part of the analysis investigates how varying patch sizes influence the computational cost of the cross-attention mechanism in CaM. The results in \cref{fig: token_size}(a) show that, as the patch size increases, the ratio of FLOPs remains relatively stable. Although the GPU usage ratio increases, even for the large patch size of $(96, 192, 192)$, the GPU usage does not exceed 20\%. This indicates that, under typical patch sizes, the cross-attention mechanism does not become a significant computational bottleneck.

The second part examines how different token sizes in the local branch affect both GPU usage and segmentation performance (measured by Dice and ASSD). In \cref{fig: token_size}(b), we test four configurations with varying token sizes: $(3,6,8,8)$, $(1,2,3,6)$, $(1,1,2,3)$, and $(1,1,1,8)$. The results show that while reducing the token size increases the number of tokens and consequently the computational cost, the performance metrics improve accordingly. This indicates that the CaM module benefits from larger token sizes, allowing for enhanced model performance without significantly compromising computational efficiency.
\begin{figure}
  \centering
    \centering
    \includegraphics[width=\linewidth]{figures/effect_tokensize.jpg}
    \caption{(a) Computational cost of cross-attention in the CaM module across different patch sizes. (b) Effect of token size in the local branch on GPU usage and performance metrics (Dice and ASSD).}
    \label{fig: token_size}
\end{figure}

\begin{table}
\centering

\begin{tabular}{lllll}
\toprule
Loss Func & Dice  & clDice & NSD   & ASSD \\
\midrule
Dice      & \underline{76.43}               & 85.13             & 88.45             & 0.74 \\
clDice    & \textbf{77.40}      & \textbf{86.00}    & \textbf{89.94}    & \textbf{0.61} \\
cbDice    & 76.31   & \underline{85.29} & \underline{88.81} & \underline{0.68} \\
\bottomrule
\end{tabular}
\caption{The performance metrics of the model using different loss functions on the TubeTK dataset.}
\label{tab: loss func}
\end{table}

\subsubsection{Analysis of Topology-aware Loss}
In this experiment, we explored the effectiveness of topology-aware loss functions, specifically clDice and cbDice, to improve the segmentation performance of our model. The results shown in the \cref{tab: loss func} reveal that using clDice significantly improved the Dice score and clDice, achieving values of 77.40\% and 86.00\%, respectively. Additionally, clDice also led to a notable increase in NSD, with a value of 89.94\%. However, the ASSD decreased to 0.61. On the other hand, the cbDice loss did not show a significant improvement in performance, with its results slightly below those of clDice but still demonstrating overall improvements compared to the original Dice loss. These findings suggest that our model can effectively adapt to topology-aware loss functions. Furthermore, the potential to further design and integrate such loss functions into the model architecture offers promising avenues for improving segmentation performance.
\begin{figure*}
  \centering
  \includegraphics[scale=0.2]{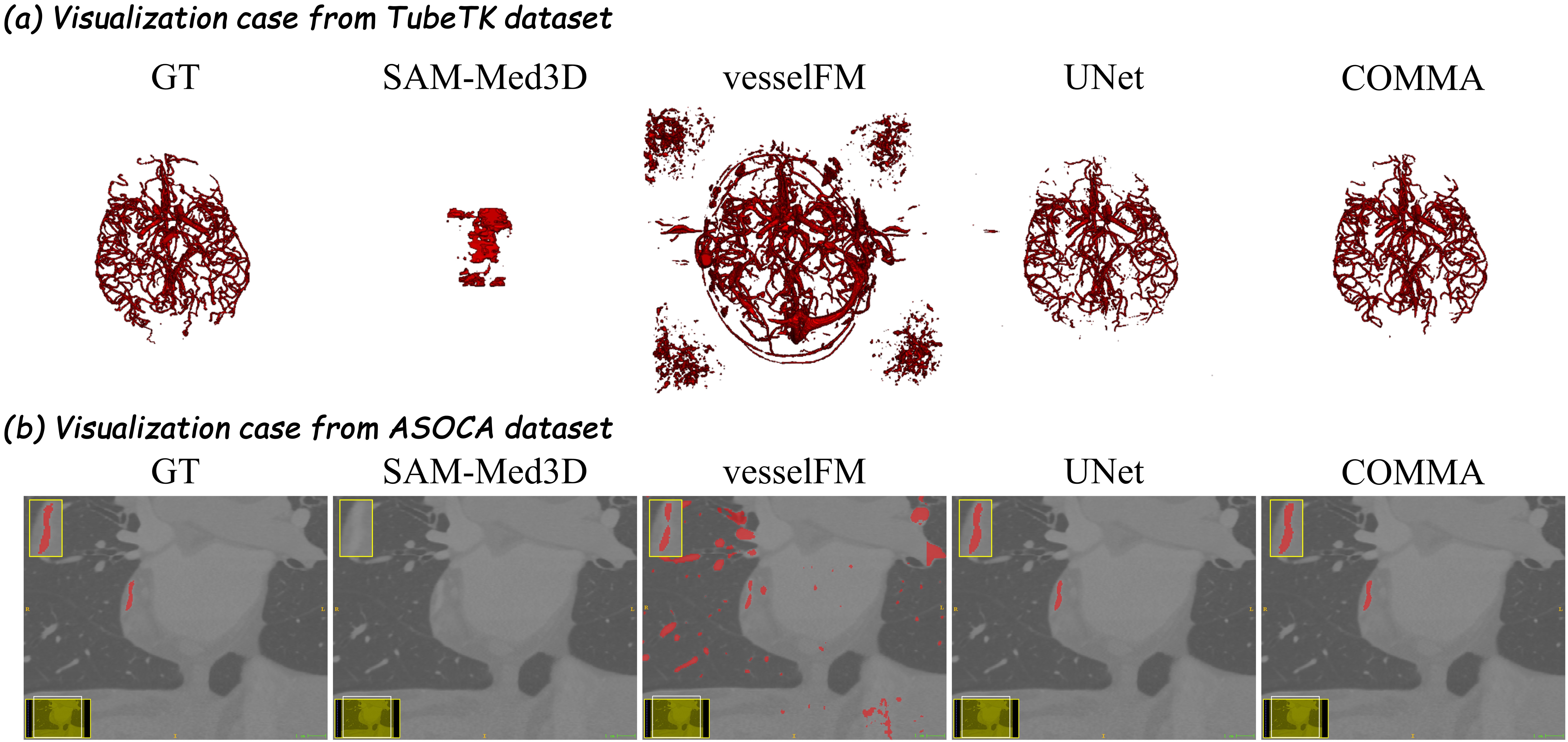}
    \caption{Visualization of segmentation results from two foundation models and two task-specific models on two datasets. (a) A 3D visualization example from the TubeTK dataset. (b) A 2D slice visualization from the ASOCA dataset.}
    \label{fig:comparison_FM}
\end{figure*}

\subsection{Analysis of Foundation Model for 3D Vessel Segmentation}
\label{sec FM analysis}
We conducted a preliminary evaluation of two foundation models for 3D medical image segmentation: SAM-Med3D \cite{wang2023sam} and vesselFM \cite{wittmann2024vesselfm}, focusing on the task of 3D vascular segmentation. The experiments were carried out on two datasets: TubeTK, which is partially included in the training set of vesselFM, and ASOCA, a dataset entirely unseen by vesselFM. This setup allowed us to assess the generalization capabilities of these foundation models in both in-domain and out-of-domain scenarios. Specifically, SAM-Med3D provides a vascular tissue point as a prompt to guide the segmentation process, while vesselFM performs predictions without any additional information, relying solely on the model’s learned features.

The quantitative results are summarized in \cref{tab:comparison_FM}, and qualitative visualizations are presented in \cref{fig:comparison_FM}. Overall, both foundation models perform poorly compared to conventional task-specific models such as UNet and the proposed COMMA. For instance, vesselFM shows relatively high performance on TubeTK (Dice: 29.18), but it significantly degrades on the unseen ASOCA dataset (Dice: 4.07). SAM-Med3D performs consistently worse on both datasets, indicating its limited adaptability to fine-scale vascular structures. 

As shown in the visualizations, SAM-Med3D struggles to segment vascular structures and fails to produce coherent vessel regions, while vesselFM can roughly identify the presence of vessels but generates numerous false positives and lacks accuracy in capturing fine vascular details. On the contrary, UNet and COMMA produce more accurate and complete segmentations. These findings suggest that although foundation models show promise, their current formulations are insufficient for precise 3D vessel segmentation tasks without further adaptation or fine-tuning.
\begin{table}
\centering
\begin{tabular}{lcccccc}
\toprule
\multicolumn{1}{c}{\textbf{Method}} & \multicolumn{3}{c}{\textbf{TubeTK}} & \multicolumn{3}{c}{\textbf{ASOCA}} \\
\cmidrule(lr){2-4} \cmidrule(lr){5-7}
 & Dice & clDice & NSD & Dice & clDice & NSD \\
\midrule
SAM-Med3D   & 00.46 &  00.62 &  01.42 &  01.30 &  00.68 &  02.34 \\
vesselFM   & 29.18 & 45.00 & 47.22 &  04.07 &  03.81 &  05.56 \\ \midrule
UNet       & 72.32 & 80.92 & 87.02 & 79.70 & 74.40 & 82.15 \\
COMMA      & 76.43 & 85.06 & 88.45 & 84.20 & 82.00 & 88.58 \\
\bottomrule
\end{tabular}
\caption{Preliminary evaluation of foundation models (SAM-Med3D and vesselFM) for 3D vascular segmentation on the TubeTK and ASOCA datasets}
\label{tab:comparison_FM}
\end{table}

\begin{table}
\centering

\begin{tabular}{lllll}
\toprule
Method     & Dice  & clDice & NSD   & ASSD \\
\midrule
UNet       & 55.08          & 70.59             & 79.14             & 1.90 \\
UNETR      & 53.98          & 72.72             & 79.09             & 1.69 \\
Swin UNETR & 53.53          & 74.13             & 80.15             & 1.61 \\
ERNet      & 55.78                  & 75.37             & 81.11             & 1.53 \\
DSCNet     & \underline{56.38}    & 75.35             & \underline{81.18}             & \underline{1.50} \\
TRA        & 54.28                 & 74.20             & 79.09             & 1.70 \\
MDNet      & 55.48          &\underline{75.84}             & 81.22             & 1.51 \\
UMamba     & 55.08          & 74.82             & 79.45             & 1.66 \\
SegMamba   & 54.85          & 75.45             & 80.71             & 1.57 \\
nnUNet     & \textbf{56.56} & 75.40             & 81.12             & 1.60 \\
COMMA      & 56.15          & \textbf{76.51}    & \textbf{82.02}    & \textbf{1.36} \\
\bottomrule
\end{tabular}
\caption{Comparison of segmentation performance across different methods, trained on the IXI dataset and tested on the TubeTK dataset.}
\label{tab:generalization}
\end{table}

\subsection{Cross-Dataset Generalization}
To investigate the cross-dataset generalization capability of our method, we conducted an experiment where we trained on the IXI dataset and tested on the TubeTK dataset without any fine-tuning or domain adaptation. The results presented in the \cref{tab:generalization} demonstrate that COMMA performs well in this cross-dataset setting, achieving notable results in metrics like clDice, NSD, and ASSD. One reason for the strong performance across datasets may be our approach of leveraging both global and local data during testing. This dual approach likely provides an advantage in capturing both broader and finer features, contributing to more accurate vessel segmentation.

\section{Discussion}
3D Vascular structures differ fundamentally from compact organs in terms of spatial distribution—they are inherently dispersed and exhibit significant spatial uncertainty. However, most existing segmentation methods, including emerging foundation models in medical image analysis, may overlook this critical characteristic. For instance, Medical-SAM \cite{ma2024segment} explicitly acknowledges its difficulty in segmenting fine-grained vessel structures. In this work, we first perform a quantitative analysis to characterize the spatial differences between vascular and compact anatomical structures. Building on this insight, we propose a task-specific dual-branch architecture with a coordinate-aware modulation mechanism to restore global spatial awareness in patch-based predictions. This design not only improves segmentation performance for dispersed structures but also offers a promising direction for developing vascular-specific foundation models.

In addition to methodological innovations, this study contributes a large-scale 3D vascular segmentation dataset. In the context of vascular structures—where annotations are particularly challenging due to their dispersed and fine-scale topology—high-quality labeled data is not only essential for task-specific model design, but also foundational for the advancement of general-purpose or foundation models. Recent foundation model vesselFM leverages synthesized annotations to enhance model performance, yet it only uses approximately 464 real cases during training \cite{wittmann2024vesselfm}. In contrast, our curated dataset of 570 manually annotated cases provides a more substantial and reliable resource. We believe that such large-scale data can facilitate both the design of vascular-specific architectures and the pretraining of generalizable models, enabling better representation learning for sparse and structurally complex anatomical targets like vessels. In addition, this dataset has the potential to promote the development of semi-supervised and self-supervised learning frameworks, where large-scale unlabeled or partially labeled data can be exploited more effectively.

In this study, the dataset was annotated by two biomedical engineering PhD students, under the guidance of two experienced clinical doctors. Initially, the students underwent training before performing a partial annotation of the data. Following the training, they independently annotated the same batch of data and discussed their results until they reached a consensus. Afterward, each student independently annotated half of the remaining data. Once the full dataset was annotated, the two clinical doctors reviewed and refined the annotations, ensuring accuracy and consistency across all samples. The annotation process utilized ITK-SNAP, and grayscale transformations were applied to enhance the visibility of the vascular structures. The annotations focused specifically on the visible intracranial vascular structures, which is the primary focus of TOF-MRA imaging for the brain. Since TOF-MRA images do not effectively display veins, they were not annotated in this dataset.

Our proposed method achieves a favorable balance between segmentation accuracy and computational efficiency. Unlike existing approaches that solely use the patch-wise inputs, COMMA simultaneously leverages both patch data and full-resolution images through a dual-branch design. Despite incorporating additional global context, COMMA maintains a lower computational burden. Specifically, its FLOPs (249.31G) are nearly 90\% lower than those of UMamba (3814.20G), and its peak GPU memory usage (8.63G) is only one-sixth of that required by DSCNet (47.00G). This efficiency stems from the adoption of the Mamba architecture in the global branch, which enables memory-efficient long-range modeling. In a direct comparison, the Mamba-based global branch consumes 7.2× less GPU memory than its Transformer-based counterpart (8.63G vs. 62.23G), while achieving slightly better segmentation performance.

One limitation of this study is that COMMA may not fully leverage unlabeled data. Due to the dispersed nature of vascular structures, annotating them is more labor-intensive than for compact organs, resulting in smaller labeled datasets. Many semi-supervised \cite{Basak_2023_CVPR, 10871927} and self-supervised \cite{shi2023progressive, 11003156} methods have been developed to exploit unlabeled data, but COMMA does not yet utilize it. Future research could explore the development of semi-supervised and self-supervised methods within the dual-branch framework, optimizing spatial location utilization to enhance vessel segmentation performance. Additionally, while inter-observer agreement was established in our annotation process, we did not conduct formal numerical evaluations to quantify inter- or intra-observer agreement. We recognize this limitation and plan to incorporate such statistical measures in future studies to better assess and quantify the reliability of the annotations.

\section{Conclusion}
In this study, we propose a Mamba-based dual-branch network for precise 3D vessel segmentation. We first quantitatively highlight the dispersed nature and spatial uncertainty of vascular structures and introduce a coordinate-based interaction strategy to enhance spatial information perception. The Mamba architecture is employed to capture long-range dependencies with relatively low computational costs. COMMA has been evaluated on six datasets, outperforming the state-of-the-art methods. Future work may explore integrating COMMA into semi-supervised or self-supervised learning paradigms to leverage unlabeled data. Additionally, the proposed coordinate-aware strategy holds potential for incorporation into foundation model design for 3D vascular segmentation.

\section*{Acknowledgments}
We would like to express our sincere gratitude to the following individuals for their invaluable contributions to this study. Our deepest appreciation goes to Haihui Jiang, M.D., Ph.D., an attending neurosurgeon with over 10 years of experience in neuroimaging analysis, from the Department of Neurosurgery, Peking University Third Hospital, Peking University, and Dr. Jiaojiao Liu from the Department of Radiology, Beijing Youan Hospital, Capital Medical University, for their expert guidance and thorough review of the annotations. Their expertise and collaboration were crucial in ensuring the quality and accuracy of the annotated dataset. We also wish to thank PhD student Wenxuan Zou from Beihang University for his dedicated work in annotating the dataset.

\bibliographystyle{IEEEtran}
\bibliography{main}

\end{document}